\documentstyle{mn2e}
\input psfig

  \makeatletter
  \ifx\CUP@mtlplain@loaded\undefined  
  \makeatother  
    
  \else
    
  \fi

\loadboldmathitalic 
\loadboldgreek      
  
\makeatletter
\ifx\CUP@mtlplain@loaded\undefined  
  \newfont\bit{cmbxti10 at 9pt}
\else
  \newfont\bit{mtbxti10 at 9pt}
\fi
\makeatother

\def\LaTeX{L\kern-.36em\raise.3ex\hbox{a}\kern-.15em
    T\kern-.1667em\lower.7ex\hbox{E}\kern-.125emX}

\newcommand{\gsim}{\mathrel{\hbox{\rlap{\lower.55ex \hbox {$\sim$}}
                   \kern-.3em \raise.4ex \hbox{$>$}}}}
\newcommand{\lsim}{\mathrel{\hbox{\rlap{\lower.55ex \hbox {$\sim$}}
                   \kern-.3em \raise.4ex \hbox{$<$}}}}

\title[The formation of close binaries]{The formation of close binary systems by dynamical interactions and orbital decay}

\author[M. R. Bate et~al.]
  {Matthew R. Bate,$^{1,2}$\thanks{E-mail: mbate@astro.ex.ac.uk}
  Ian A. Bonnell,$^3$
  and Volker Bromm.$^{2,4}$\\
  $^1$School of Physics, University of Exeter, Stocker Road,
    Exeter EX4 4QL \\
  $^2$Institute of Astronomy, University of Cambridge, Madingley Road,
    Cambridge CB3 0HA \\
  $^3$School of Physics and Astronomy, University of St Andrews, North Haugh, St Andrews, Fife, KY16 9SS \\
  $^4$Harvard-Smithsonian Center for Astrophysics, 60 Garden Street, Cambridge,
MA 02138, U.S.A.
}

\date{Accepted for publication in MNRAS}

\begin{document}

\maketitle

\begin{abstract}
  We present results from the first hydrodynamical star formation
  calculation to demonstrate that close binary stellar systems 
  (separations $\lsim 10$ AU) need not be formed 
  directly by fragmentation.  Instead, a high frequency of close 
  binaries can be produced through a combination of dynamical 
  interactions in unstable multiple systems and the orbital decay 
  of initially wider binaries.  Orbital decay may occur due to 
  gas accretion and/or the interaction of a binary with its circumbinary 
  disc.  These three mechanisms avoid the problems associated with the 
  fragmentation of optically-thick gas to form close systems directly.  
  They also result in a preference for close binaries to have roughly
  equal-mass components because dynamical exchange interactions 
  and the accretion of gas with high specific angular momentum 
  drive mass ratios towards unity.  Furthermore, due to the importance of 
  dynamical interactions, we find that stars with greater masses 
  ought to have a higher frequency of close companions, and 
  that many close binaries ought to have wide companions.  
  These properties are in good agreement with the results of 
  observational surveys.
\end{abstract}


\section{Introduction}

The process of star formation preferentially produces binary stellar 
systems (e.g.\ Duquennoy \& Mayor 1991).  The favoured mechanism
for explaining this high frequency of binaries is the collapse and 
fragmentation of molecular cloud cores (e.g.\ Boss \& Bodenheimer 1979; 
Boss 1986; Bonnell et al.\ 1991; Nelson \& Papaloizou 1993; 
Burkert \& Bodenheimer 1993; Bate, Bonnell \& Price 1995).  
However, while fragmentation can readily create wide binary systems 
(separations $\gsim 10$ AU), there are severe
difficulties with fragmentation producing close binaries
directly.  This is a significant deficiency since approximately 20\% of
solar-type stars have main-sequence companions that orbit closer than 10 AU
(Duquennoy \& Mayor 1991), and the frequency of massive spectroscopic
binaries appears to be even higher 
(Garmany, Conti \& Massey 1980; Abt et al.\ 1990; Morrell \& Levato 1991;
Mason et al.\ 1998).

As a molecular cloud core begins to collapse, the formation of
wide binaries through fragmentation 
is possible because the gas easily radiates away the
gavitational potential energy that is released.  The gas remains
approximately isothermal and, thus, the Jeans mass decreases with
density as $\rho^{-1/2}$.  However, at densities of $\gsim 10^{-13}$
g~cm$^{-3}$ or $n({\rm H}_2)\gsim 10^{10}$ cm$^{-3}$ 
(Larson 1969; Masunaga \& Inutsuka 2000) 
the rate of heating from dynamical collapse exceeds the rate at 
which the gas can cool. The gas heats up, and the Jeans mass 
begins to increase so that a Jeans-unstable region of gas becomes 
Jeans-stable.  This results in the formation of a pressure-supported 
fragment with a mass of several Jupiter-masses and a radius of 
$\approx 5$ AU (Larson 1969).  Fragmentation on smaller scales is 
inhibited by thermal pressure.  Therefore, initial binary separations
must be $\gsim 10$ AU.  The formation of such pressure-supported fragments
is frequently refered to as the opacity limit for fragmentation
(Low \& Lynden-Bell 1976; Rees 1976) and may set a lower limit to
the mass of brown dwarfs (Boss 1988; Bate, Bonnell \& Bromm 2002).  

A possibility for fragmentation at higher densities (hence on
smaller length scales) exists when the pressure-supported fragment
has accreted enough material for its central temperature to exceed
2000 K.  At this temperature, molecular hydrogen begins to dissociate,
which provides a way for the release of gravitational energy to be absorbed
without significantly increasing the temperature of the gas.  Thus, a
nearly isothermal second collapse occurs within the fragment that
ultimately results in the formation of a stellar core with radius
$\approx 1 R_\odot$ \cite{Larson1969}.
Several studies have investigated the possibility that fragmentation 
during this second collapse forms close binary systems directly 
(Boss 1989; Bonnell \& Bate 1994; Bate 1998, 2002).  Boss \shortcite{Boss1989}
found that fragmentation was possible during this second collapse, but that 
the fragments spiralled together due to gravitational torques and 
did not survive.  
Bonnell \& Bate \shortcite{BonBat1994} found that fragmentation
to form close binaries and multiple systems could occur in a disc 
that forms around the stellar core.  However, both these studies
began with somewhat arbitrary initial conditions for the pressure-supported
fragment.  Bate \shortcite{Bate1998} performed the first three-dimensional
calculations to follow the collapse of a molecular cloud core through the
formation of the pressure-supported fragment, the second collapse,
and the formation of the stellar core and its surrounding disc.
In these and subsequent calculations (Bate 2002), Bate found that the second
collapse did not result in sub-fragmentation due to the high thermal
pressure and angular momentum transport via gravitational torques.

In this paper, we present results from the first hydrodynamical 
star formation calculation to produce dozens of stars and brown
dwarfs while simultaneously resolving beyond the opacity limit for
fragmentation.  Despite the fact that no close binaries 
(separations $\lsim 10$ AU) are formed by direct fragmentation, we find
that the 
calculation eventually produces several close binary systems through
a combination of dynamical interactions in multiple systems, and 
the orbital decay of wide binaries via gas accretion and their interactions 
with circumbinary discs.

The paper is structured as follows. In section 2, we briefly
describe the numerical method and the initial conditions for
our calculation.  In section 3, we present results from our
calculation and compare them with observations.  Finally, in 
section 4, we give our conclusions.

\section{Computational method and initial conditions}

The calculation presented here was performed using a three-dimensional, 
smoothed particle hydrodynamics (SPH) code.  The SPH code is 
based on a version originally developed by Benz (Benz 1990; 
Benz et al.~1990).  The smoothing lengths of particles are variable in 
time and space, subject to the constraint that the number of 
neighbours for each particle must remain approximately constant 
at $N_{\rm neigh}=50$.  We use the standard form of artificial viscosity 
(Monaghan \& Gingold 1983) with 
strength parameters $\alpha_{\rm_v}=1$ and $\beta_{\rm v}=2$.
Further details can be found in Bate et al.\ \shortcite{BatBonPri1995}.  
The code has been parallelised by M. Bate using OpenMP.

\subsection{Opacity limit for fragmentation and the equation of state}

To model the opacity limit for fragmentation without performing 
full radiative transfer, we use a barotropic equation of state for 
the thermal pressure of the gas $p = K \rho^{\eta}$,
where $K$ is a measure of the entropy of the gas.  The value of 
the effective polytropic exponent $\eta$,
varies with density as
\begin{equation}\label{eta}
\eta = \cases{\begin{array}{rl}
1, & \rho \leq 10^{-13}~ {\rm g~cm}^{-3}, \cr
7/5, & \rho > 10^{-13}~ {\rm g~cm}^{-3}. \cr
\end{array}}
\end{equation}
We take the mean molecular weight of the gas to be $\mu = 2.46$.
The value of $K$ is defined such that when the gas is 
isothermal $K=c_{\rm s}^2$, with the sound speed
$c_{\rm s} = 1.84 \times 10^4$ cm~s$^{-1}$ at 10 K,
and the pressure is continuous when the value of $\eta$ changes.

This equation of state reproduces the temperature-density relation
of molecular gas during spherically-symmetric collapse 
(as calculated with frequency-dependent radiative transfer)
to an accuracy of better than 20\% in the non-isothermal 
regime up to densities of 
$10^{-8}~{\rm g~cm}^{-3}$ \cite{MasInu2000}.  
Thus, our equation of state should model 
collapsing regions well, but may not model the equation of 
state in protostellar discs particularly accurately due to
their departure from spherical symmetry.

\subsection{Sink particles}

The opacity limit results in the formation of distinct pressure-supported
fragments in the calculation.  As these fragments accrete, their
central density increases, and it becomes computationally impractical 
to follow their internal evolution until they undergo the second
collapse to form stellar cores because of the short dynamical time-scales
involved.  Therefore, when the central density of a pressure-supported 
fragment exceeds $\rho_{\rm s} = 10^{-11}~{\rm g~cm}^{-3}$, we insert a sink 
particle into the calculation (Bate et al.\ 1995).  The gas within
radius $r_{\rm acc}=5$ AU of the centre of the fragment 
(i.e.\ the location of the SPH particle with the highest density)
is replaced by a point mass with the same mass and momentum.
Any gas that later falls within this radius is accreted by the 
point mass if it is bound and its specific angular momentum is 
less than that required to form a circular orbit at radius 
$r_{\rm acc}$ from the sink particle.  Thus, gaseous discs 
around sink particles can only be resolved if they have 
radii $\gsim 10$ AU.  Sink particles interact with the gas 
only via gravity and accretion.

Since all sink particles are created from pressure-supported 
fragments, their initial masses are $\approx 10$ Jupiter-masses
(M$_{\rm J}$), as 
given by the opacity limit for fragmentation \cite{Boss1988}.
Subsequently, they may accrete large amounts of material 
to become higher-mass brown dwarfs ($\lsim 75$ M$_{\rm J}$) or 
stars ($\gsim 75$ M$_{\rm J}$), but all the stars and brown dwarfs
begin as these low-mass pressure-supported fragments.

The gravitational acceleration between two sink particles is Newtonian 
for $r \geq 4$ AU, but is softened within this radius using spline 
softening \cite{Benz1990}.  The maximum acceleration occurs at a 
distance of $\approx 1$ AU; therefore, this is the minimum 
separation that a binary can have even if, in reality, the binary's
orbit would have been hardened.

Replacing the pressure-supported fragments with sink particles 
is necessary in order to perform the calculation.  However, it is 
not without a degree of risk.  If it were possible to follow the 
fragments all the way to stellar densities (as done by Bate 1998)
and still follow the evolution of the large-scale cloud over its 
dynamical time-scale, we might find that a few of the objects that we 
replace with sink particles merge together or are
disrupted by dynamical interactions.  We have tried to minimise the 
degree to which this may occur by insisting that the central
density of the pressure-supported fragments exceeds $\rho_{\rm s}$
before a sink particle is created.  This is two orders of magnitude 
higher than the density at which the gas is heated and ensures that
the fragment is self-gravitating, centrally-condensed and, in practice, 
roughly spherical before it 
is replaced by a sink particle.  In theory, it would be possible 
for a long collapsing filament to exceed this density over a large
distance, thus making the creation of one or more sink 
particles ambiguous.  However, the structure of the collapsing 
gas that results 
from the turbulence prohibits this from occuring; no long roughly 
uniform-density filaments with densities $\approx \rho_{\rm s}$
form during the calculation.  Furthermore, each pressure-supported 
fragment must undergo a period of accretion before its central density
exceeds $\rho_{\rm s}$ and it is replaced by a sink particle.  
For example, it is common in the calculation
to be able to follow a pressure-supported fragment that forms via 
gravitational instability in a disc for roughly half an orbital 
period before it is replaced.  Thus, the fragments do have some time 
in which they may be disrupted 
or merge.  Only occasionally during the calculation are low-mass 
pressure-supported fragments disrupted; most are eventually 
replaced by sink particles.

\begin{table*}
\begin{tabular}{lccccl}\hline
Binary & M$_1$ & M$_2$ & q & a & Notes and the end of the calculation\\
 & M$_\odot$ & M$_\odot$  & & AU  \\ \hline
 3,10 & 0.73 & 0.41 & 0.56 & 1.1* & In hierarchical triple, 0.083 M$_\odot$ at 28 AU; circumtriple disc; member of a bound group\\
 7,8  & 0.53 & 0.24 & 0.44 & 2.0* & Ejected from cloud \\
20,22 & 0.35 & 0.11 & 0.33 & 2.2* & In hierarchical triple, 0.23 M$_\odot$ at 28 AU; circumtriple disc; member of a bound group\\
26,40 & 0.13 & 0.039 & 0.29 & 6.7~ & Circumbinary disc; member of septuple system with (39,41), ((45,38),43)\\
39,41 & 0.070 & 0.047 & 0.67 & 5.7~ & Circumbinary disc; member of septuple system with (26,40), ((45,38),43)\\
42,44 & 0.10 & 0.095 & 0.93 & 2.6* & In unstable quintuple system; circumquadruple disc\\
45,38 & 0.083 & 0.079 & 0.96 & 8.8~ & In hierarchical triple, 0.022 M$_\odot$ at 90 AU; circumtriple disc; member of septuple system\\
\hline
\end{tabular}
\caption{\label{table1} The properties of the 7 close binary systems at the end of the calculation.  The numbers identifying the individual stars or brown dwarfs from which the binary systems are composed are allocated in order of their formation.  We list the masses of the two components M$_1$ and M$_2$, the mass ratio $q$, and the semi-major axis $a$.  Note that the mass ratios tend to be high, with no values less than $q=0.29$ and most greater than $q=1/2$.  Asterisks indicate when the semi-major axis is less than the gravitational softening length.  If gravitational softening had not been used, these systems may have been hardened further.  }
\end{table*}

\subsection{Initial conditions}

The initial conditions consist of a large-scale, turbulent 
molecular cloud.  The cloud is spherical and
uniform in density with a mass of 50 M$_\odot$ and
a diameter of $0.375$ pc (77400 AU).  At the temperature of 10 K,
the mean thermal Jeans mass is 1 M$_\odot$ 
(i.e.\ the cloud contains 50 thermal Jeans masses).
The free-fall time of the cloud is $t_{\rm ff}=6.0\times 10^{12}$ s
or $1.90\times 10^5$ years.

Although the cloud is uniform in density, we impose an initial 
supersonic turbulent velocity field on it in the same manner
as Ostriker, Stone \& Gammie \shortcite{OstStoGam2001}.  We generate a
divergence-free random Gaussian velocity field with a power spectrum 
$P(k) \propto k^{-4}$, where $k$ is the wavenumber.  
In three dimensions, this results in a 
velocity dispersion that varies with distance, $\lambda$, 
as $\sigma(\lambda) \propto \lambda^{1/2}$ in agreement with the 
observed Larson scaling relations for molecular clouds \cite{Larson1981}.
The velocity field is normalised so that the kinetic energy of the 
turbulence equals the magnitude of the gravitational potential energy of 
the cloud.  The initial root-mean-square Mach number of the 
turbulence is ${\cal M}=6.4$.

\subsection{Resolution}

The local Jeans mass must be resolved throughout 
the calculation to model fragmentation correctly (Bate \& Burkert 1997;
Truelove et al.\ 1997; Whitworth 1998; Boss et al.\ 2000).  Bate \& Burkert 
\shortcite{BatBur1997} found that this requires 
$\gsim 2 N_{\rm neigh}$ SPH particles per Jeans mass; 
$N_{\rm neigh}$ is insufficient.
We have repeated their calculation using different numbers of
particles and find that $1.5 N_{\rm neigh}=75$
particles is also sufficient
to resolve fragmentation (Bate, Bonnell \& Bromm, in preparation).
The minimum Jeans mass in the calculation presented here occurs 
at the maximum density during the isothermal phase of the 
collapse, $\rho = 10^{-13}$ g~cm$^{-3}$, 
and is $\approx 0.0011$ M$_\odot$ (1.1 M$_{\rm J}$).  Thus, we
use $3.5 \times 10^6$ particles to model the 50-M$_\odot$ cloud.
In fact, a gas clump with the above mass and density could not 
collapse because as soon as it was compressed it would heat up 
(equation \ref{eta}) and would no longer contain a Jeans mass.  
Therefore, in practice, any collapsing gas clump in 
the simulation contains many more SPH particles than the
`minimum' Jeans mass above.
This SPH calculation is one of the largest ever performed.
It required approximately 95000 CPU hours on the SGI Origin 3800 
of the United Kingdom Astrophysical Fluids Facility (UKAFF).

\section{Results}
\label{evolution}

\subsection{Evolution of the cloud}

The hydrodynamical evolution of the cloud produces shocks which
decrease the turbulent kinetic energy initially supporting
the cloud.  In parts of the cloud, gravity begins to dominate 
and dense self-gravitating cores form and collapse.  These dense cores 
are the sites where the formation of stars and brown dwarfs occurs.
Although the cloud initially contains more than enough turbulent 
energy to support itself against gravity, this turbulence decays 
on the dynamical
time-scale of the cloud and star formation begins after just one 
global free-fall time at $t=1.04 t_{\rm ff}$ 
(i.e.\ $t=1.97\times 10^5$ yrs).  This rapid decay of the turbulence
is consistent with other numerical studies of turbulence in molecular
clouds (e.g.\ MacLow et al.\ 1998; Stone, Ostriker \& Gammie 1998; 
Ostriker et al.\ 2001).  We followed the calculation
for $\approx 69000$ years after the star formation began 
until $t=1.40 t_{\rm ff}$ ($t=2.66\times 10^5$ yrs).  At this point 
we had exhausted our allocation of computer time and the calculation
was stopped.

\begin{figure*}
\centerline{\psfig{figure=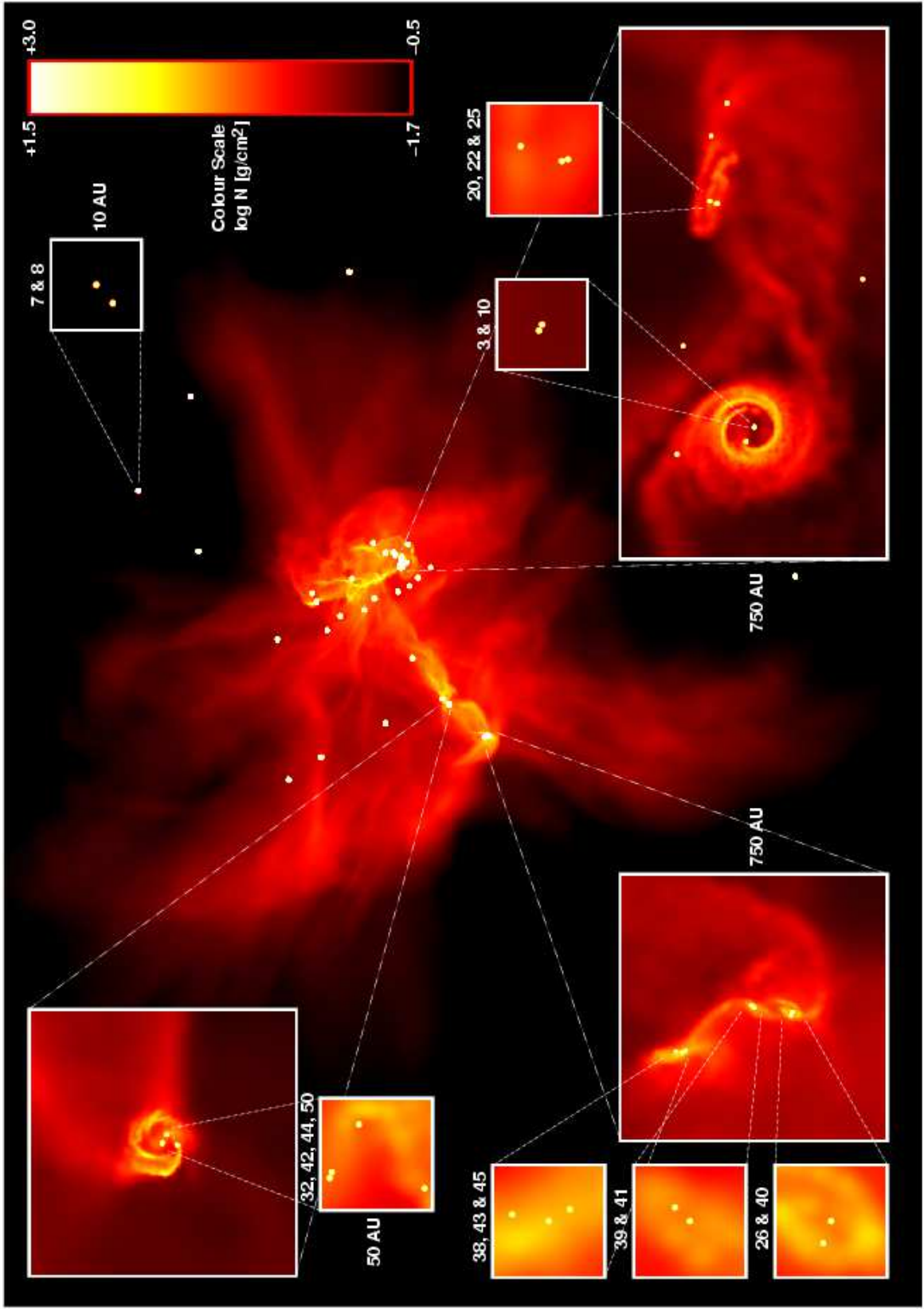,width=15.46truecm,height=22truecm,rwidth=15.46truecm,rheight=22.0truecm}}
\caption{\label{pic0} The locations and states of the 7 close binary systems when the simulation was stopped ($t=1.40 t_{\rm ff}=2.66\times 10^5$ yr).  Most of the close binaries remain members of multiple systems and are surrounded by discs, although one (top right) has been ejected from the cloud and does not have a resolved disc.  Star formation occurs in three dense cores.  The colour scales show the logarithm of column density with $-1.7 < \log N < 1.5$ for the main picture and $-0.5 < \log N < 3.0$ for each of the inserts.  The main picture is $\approx 80000$ AU from top to bottom.  The dimensions of the inserts are given in AU (inserts of equal size have equal dimensions).  The numbers identifying the objects pictured in the 10-AU and 50-AU inserts are given above each insert to allow comparison with Table 1 and Figure 2.}
\end{figure*}

Figure \ref{pic0} illustrates the state of the cloud at the 
end of the calculation.  The star formation occurs in three 
main dense cores within the cloud, resulting in three stellar 
groups.  These groups are composed of single, binary and higher-order
systems that were produced via a combination of the fragmentation of 
collapsing gas filaments (whose collapse is halted when the gas 
heats up at high densities; Inutsuka \& Miyama 1992), 
the fragmentation of massive
circumstellar discs (e.g.\ Bonnell\ 1994; Whitworth et al.\ 1995;
Burkert, Bate, \& Bodenheimer 1997), and star-disc 
capture (Larson 1990; Clarke \& Pringle 1991a,b).  
At any particular time, the largest of the
stellar groups contains no more than $\approx 20$ objects.  Thus,
the groups dissolve quickly.  In fact, the time-scales for star formation 
within the dense cores and the dissolution of the groups 
are both $\approx 2\times 10^4$ years.  Thus, the stellar groups undergo 
chaotic evolution 
with stars and brown dwarfs being ejected from the cloud even as new
objects are forming.  If the calculation were followed until well after
star formation had ceased, all of the groups would dissolve on time-scales
of a few tens of thousands of years.

When the calculation is stopped,
the cloud has produced 23 stars and 18 brown dwarfs.
An additional 9 objects have substellar masses, but are 
still accreting.  Three of these would probably end up with 
substellar masses if the calculation were continued, but the 
other six are likely to become stars.  The formation mechanism 
of the brown dwarfs in this calculation has been discussed
by Bate et al.\ \shortcite{BatBonBro2002}.  The evolution 
of the cloud and the properties of the stars and brown dwarfs 
will be discussed in detail in subsequent papers.  In this paper, 
we concentrate on the mechanism by which the close binary systems form.

\begin{figure*}
\centerline{\psfig{figure=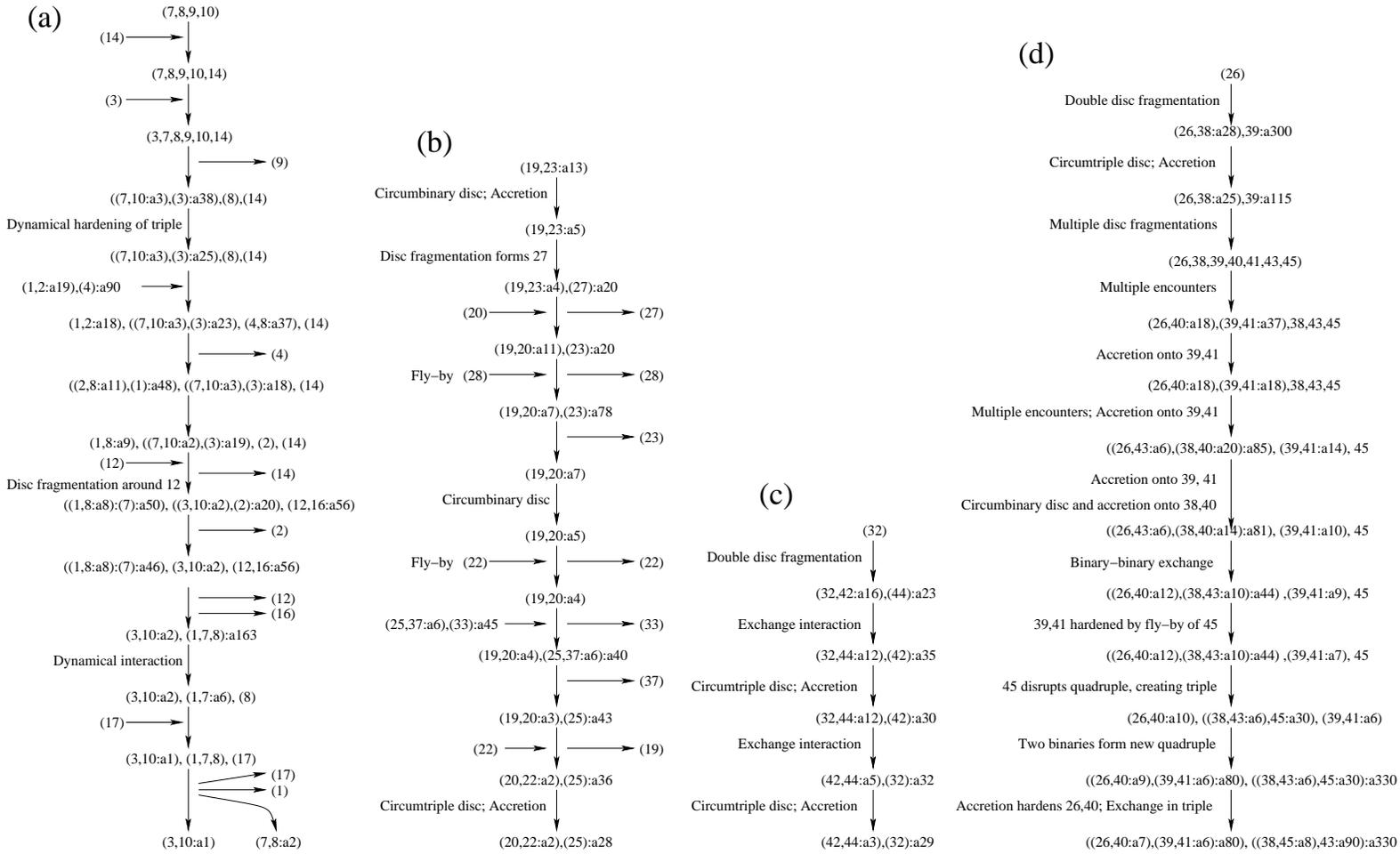,width=16.23truecm,height=21truecm,rwidth=16.23truecm,rheight=21.0truecm}}
\caption{\label{pic1} Diagrams showing the sequences of events leading to the formation of the 7 close binary systems in the course of the calculation.  Time increases in each sequence from top to bottom.  Causes of evolution (such as another object falling into a system, or the interaction of the system with a circumbinary disc) and ejected objects (if any) are given to the left and right of each sequence, respectively.  Groups of objects and hierachical multiple systems are indicated by the use of parentheses.  For example, (7,8,9,10,14) indicates 5 objects in a group with no obvious hierarchy.  Another example is ((7,10:a3),(3):a38),(8),(14) which indicates a binary composed of objects 7 and 10 with a semi-major axis of 3 AU with a companion (object 3) in an orbit with a 38-AU semi-major axis.  In turn, this hierarchical triple system ((7,10:a3),(3):a38) is part of a larger-scale group consisting of the triple system and objects 8 and 14.}
\end{figure*}

\begin{figure}
\centerline{\psfig{figure=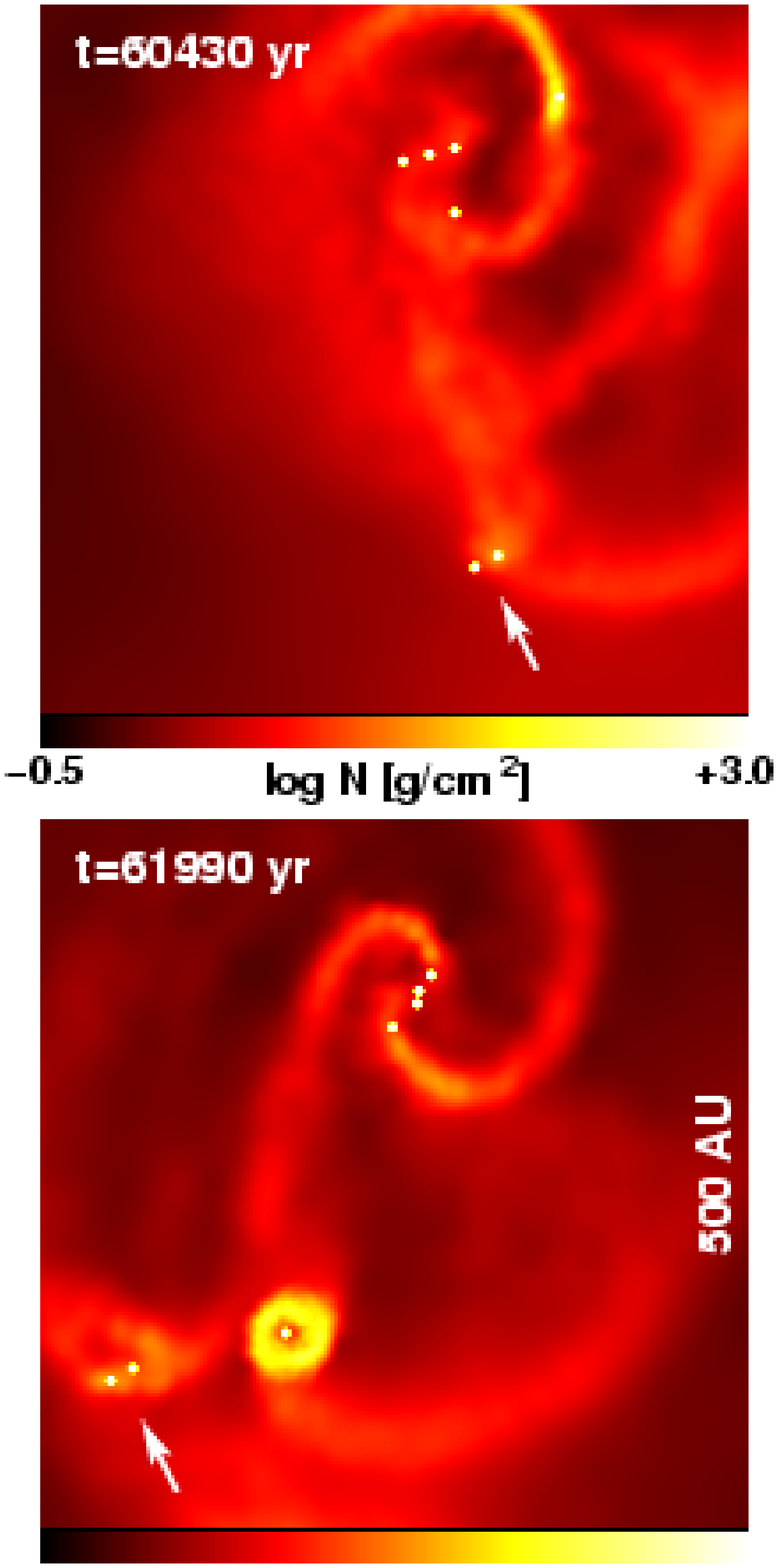,width=3.91truecm,height=7.9truecm,rwidth=3.91truecm,rheight=7.9truecm}\psfig{figure=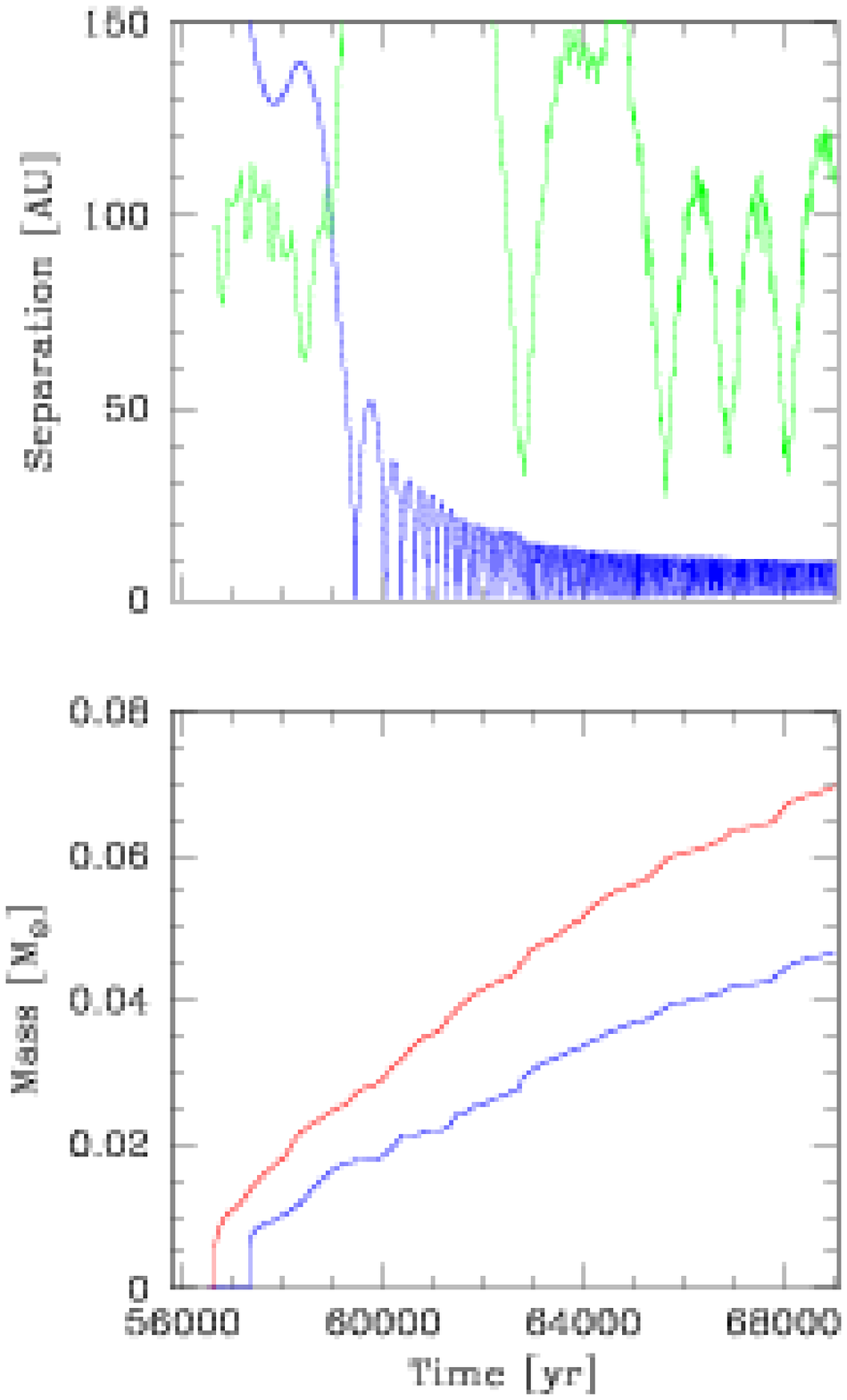,width=8.0truecm,height=8truecm,rwidth=4.8truecm,rheight=8.0truecm}}
\caption{\label{picacc} As an example of the way in which accretion can harden the orbit of a binary, we present the evolution of brown dwarfs 39 and 41 versus time (see also Figure 2d, steps 5--7).  Left panels: column density $\log N$ at two times during the evolution.  The binary, consisting of objects 39 \& 41, is indicated by the arrow.  Top-right panel: the separation of brown dwarfs 39 \& 41 (blue) and the distance between brown dwarf 39 and the closest object apart from 41 (green).  Bottom-right panel: the masses of 39 (red) and 41 (blue).  Dynamical interactions in an unstable multiple system force 39 and 41 to form an eccentric binary system with a semi-major axis of $\approx 37$ AU at $t\approx 59000$ yrs.  Over the next 5000 years, accretion reduces the semi-major axis of this binary by nearly an order of magnitude.  Dynamical interactions have only a minimal effect on the binary's separation during this period (as shown by the green line).  Note that object 39 has a mass of 0.070 M$\odot$ at the end of the calculation (Table 1) and would probably enter the stellar-mass regime if the calculation were continued.  The final mass of object 41 is unclear.   Time is given in years after the onset of star formation (star formation begins at $t=1.04 t_{\rm ff}=1.97\times 10^5$ yrs).}
\end{figure}

\begin{figure}
\centerline{\psfig{figure=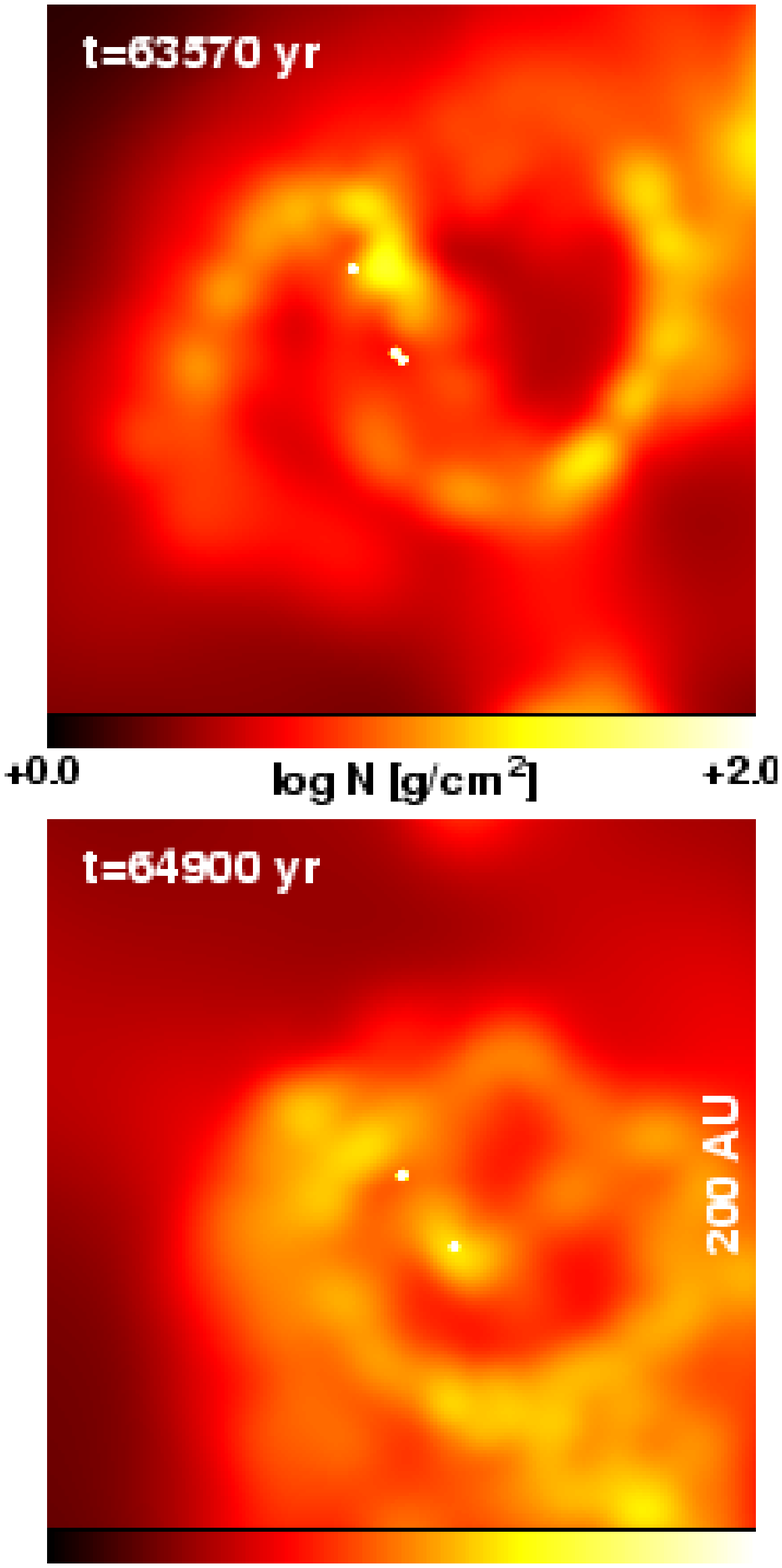,width=3.91truecm,height=7.9truecm,rwidth=3.91truecm,rheight=7.9truecm}\psfig{figure=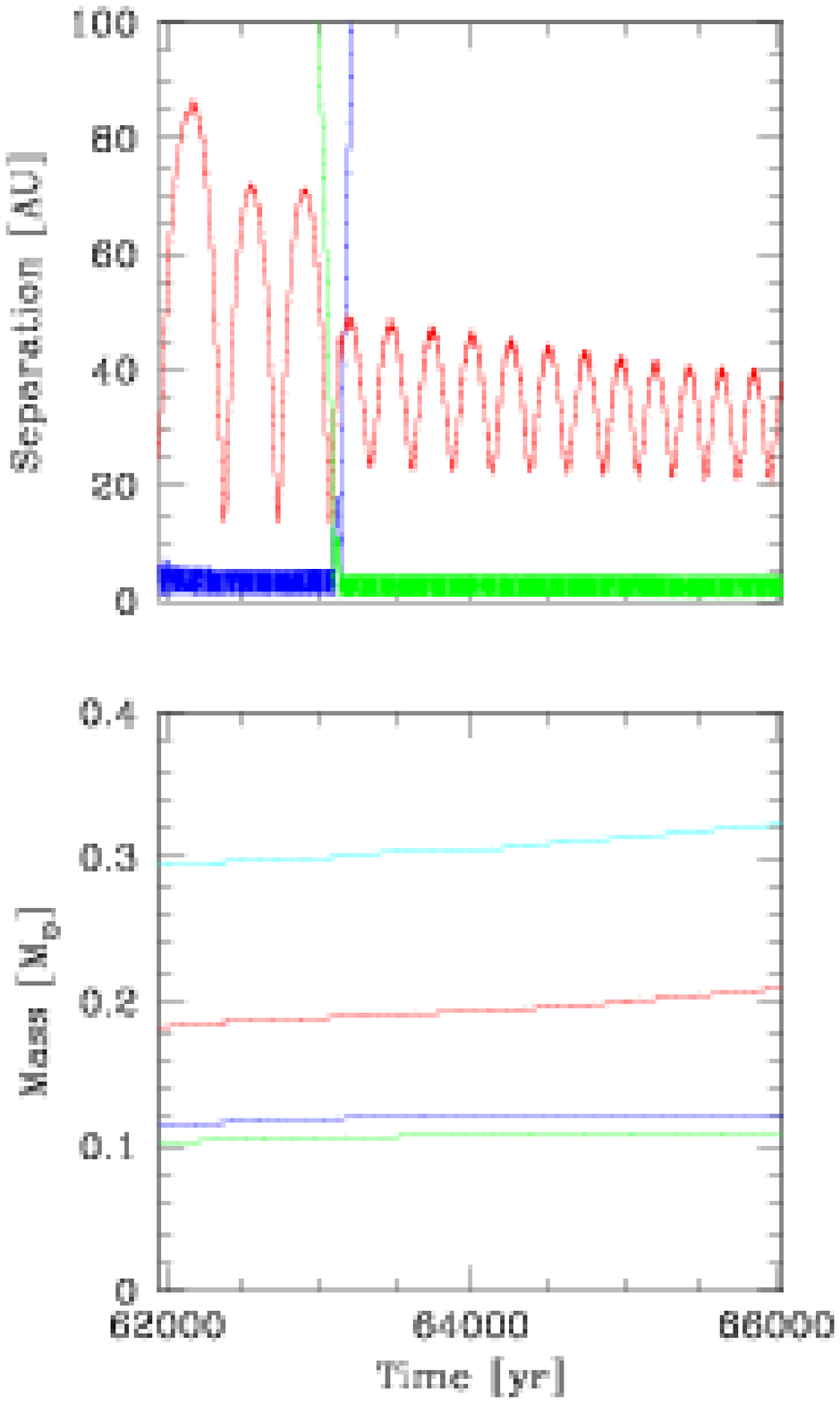,width=8.0truecm,height=8truecm,rwidth=4.8truecm,rheight=8.0truecm}}
\caption{\label{pictripacc} An example of the way in which gravitational torques and accretion from a circumtriple disc can alter the orbit of a triple system.  Left panels: column density $\log N$ at two times during the evolution.  Top-right panel: the separations of stars 19 (blue), 22 (green), and 25 (red) from star 20.  Bottom-right panel: the masses of stars 19 (blue), 20 (cyan), 22 (green), and 25 (red).  A triple system consisting of a close binary (stars 19 and 20) and a wide companion (star 25) undergoes an exchange interaction at $t\approx 63000$ yr in which star 19 is ejected and replaced by star 22.  Subsequently, the system evolves only through its interaction with the circumtriple disc which decreases the semi-major axis of star 25's orbit around the close binary (stars 20 and 22) from 36 to 28 AU (see also Figure 2b, steps 10 and 11).  Note that the circumtriple disc contains a significant fraction of the system's mass and is subject to gravitational instabilites, as demonstrated by the spiral density waves.  Time is given in years after the onset of star formation.}
\end{figure}

\begin{figure}
\centerline{\psfig{figure=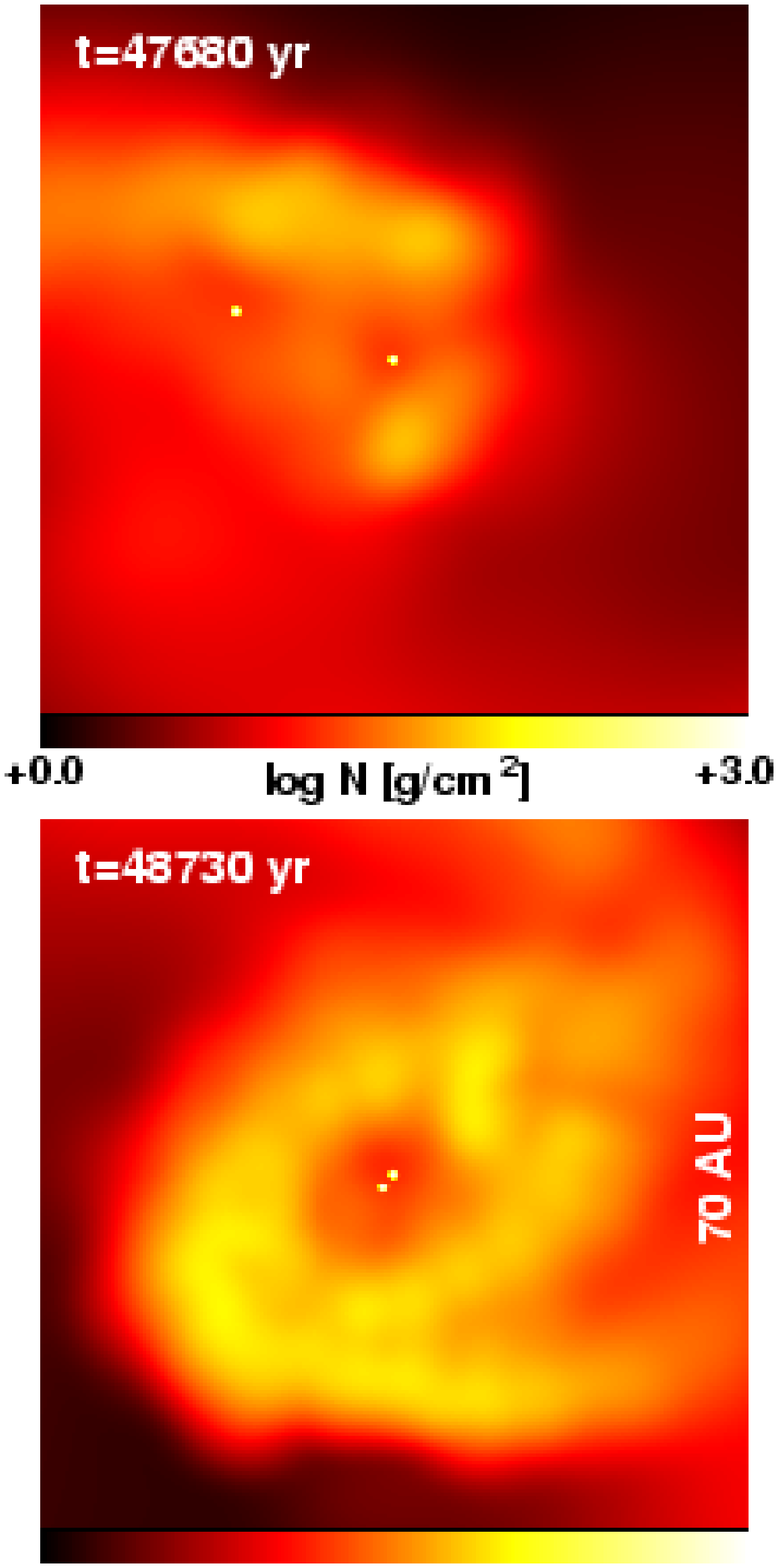,width=3.91truecm,height=7.9truecm,rwidth=3.91truecm,rheight=7.9truecm}\psfig{figure=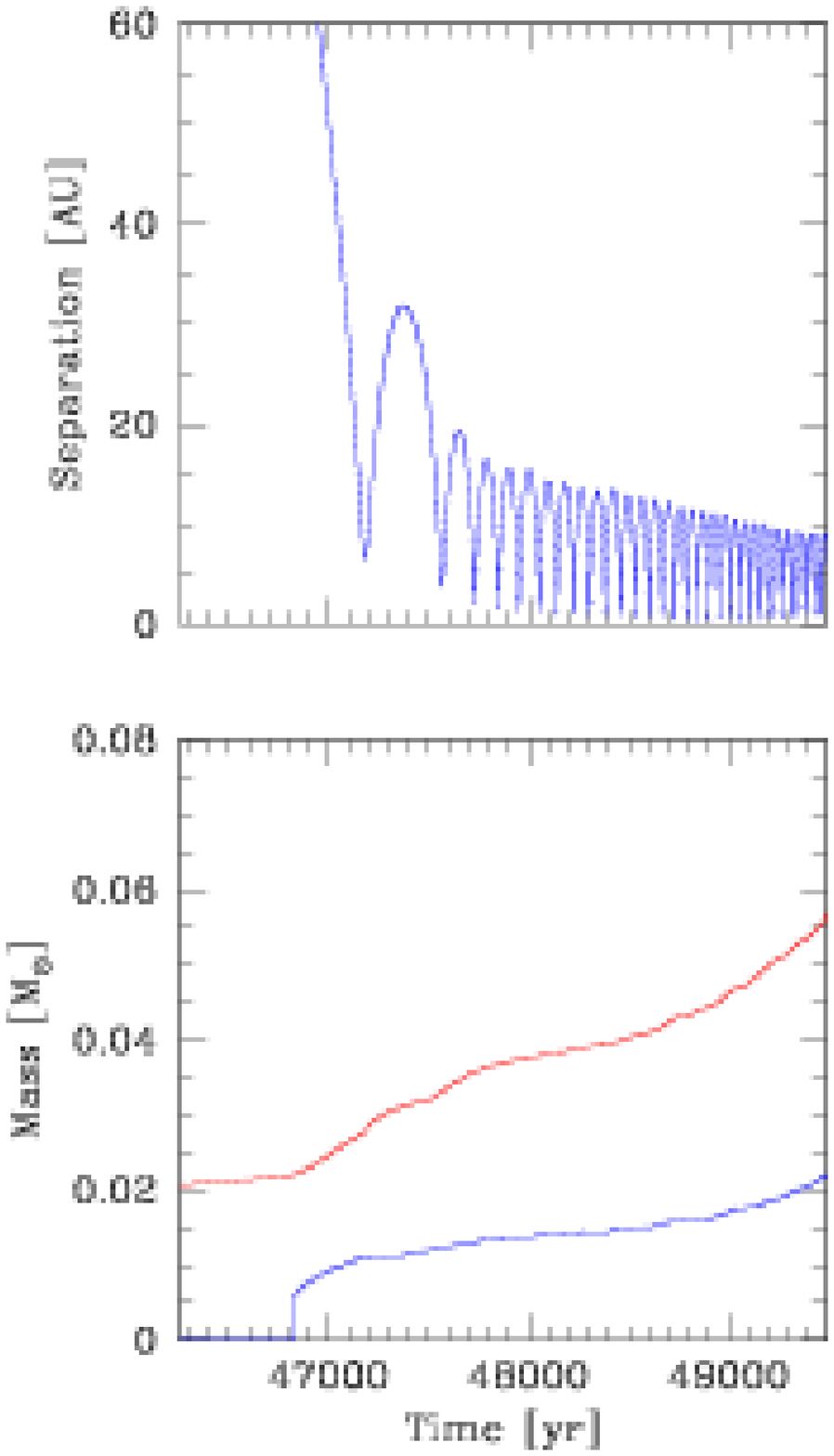,width=8.0truecm,height=8truecm,rwidth=4.8truecm,rheight=8.0truecm}}
\caption{\label{picacccb} An example of the way in which gravitational torques and accretion from a circumbinary disc can harden the orbit of a binary (see also Figure 2b, step 1).  Left panels: column density $\log N$ at two times during the evolution.  Right-panels: the evolution of the separation and masses of brown dwarfs 19 (red) and 23 (blue) versus time.  Time is given in years after the onset of star formation.}
\end{figure}

\begin{figure}
\centerline{\psfig{figure=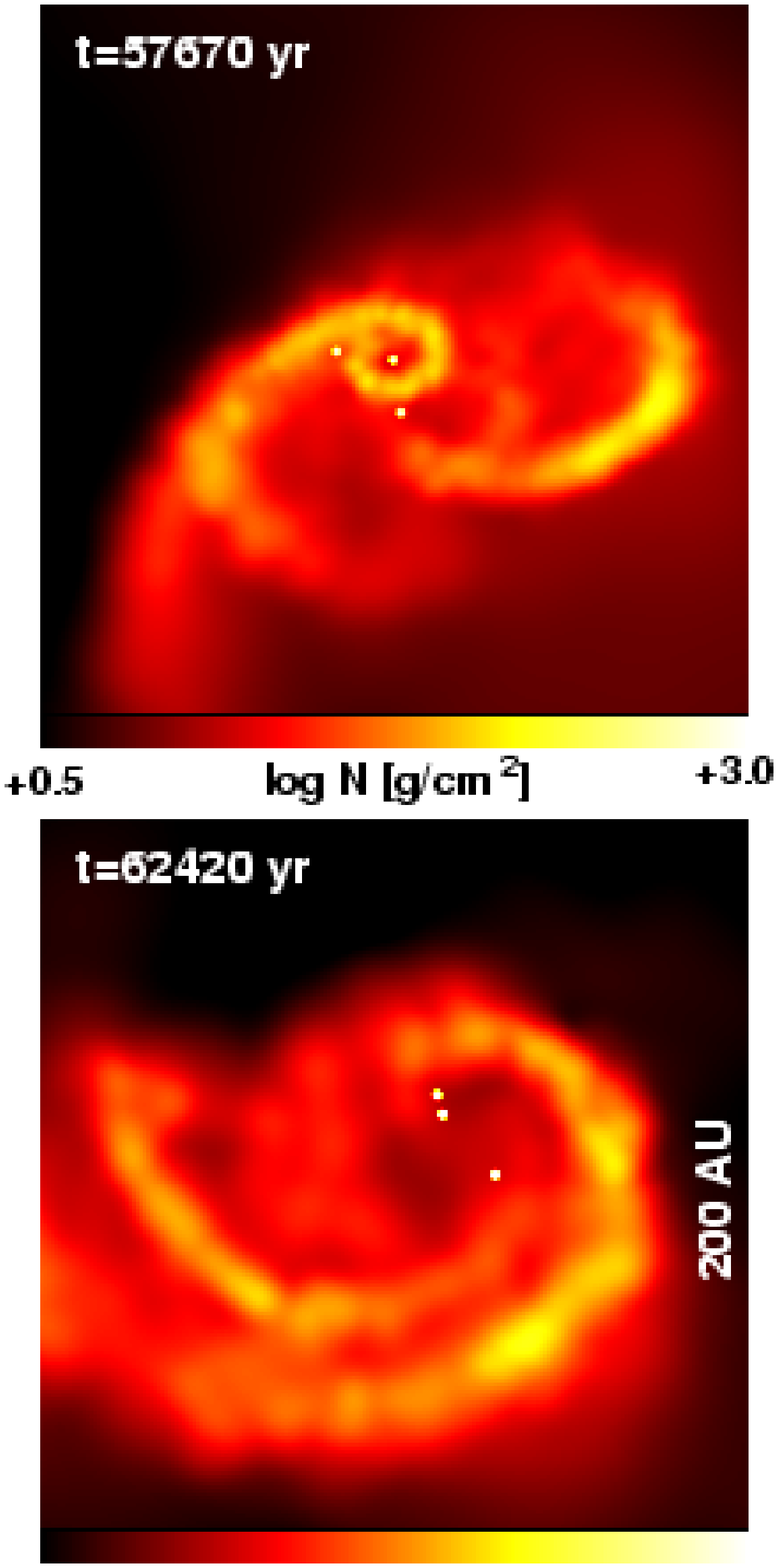,width=3.91truecm,height=7.9truecm,rwidth=3.91truecm,rheight=7.9truecm}\psfig{figure=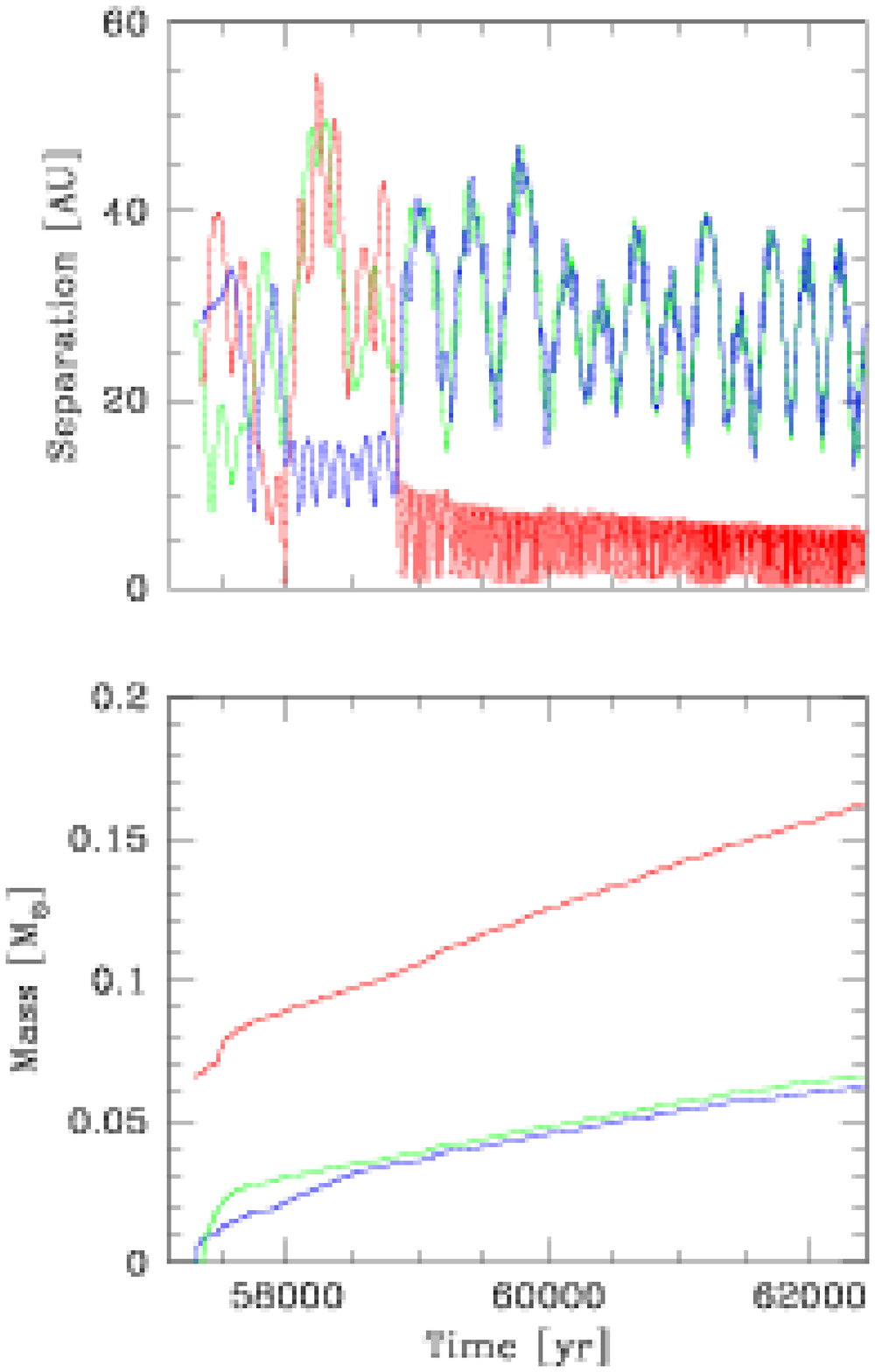,width=8.0truecm,height=8truecm,rwidth=4.8truecm,rheight=8.0truecm}}
\caption{\label{picdynacc} An example of how a combination of dynamical interactions, the interaction of a triple system with a disc, and accretion can produce close binaries (see also Figure 2c).  Left panels: column density $\log N$ at two times during the evolution.  Bottom-right panel: the masses of objects 32 (red), 42 (green), and 44 (blue) versus time.  Top-right panel: the separations of stars 32 \& 42 (blue), 32 \& 44 (green), and 42 \& 44 (red).  An unstable triple system consisting of a 16-AU binary (stars 32 \& 42) and a wide companion (star 44) undergoes an exchange interaction at $t\approx 58000$ yr to produce a hierarchical triple consisting of a 12-AU binary (stars 32 \& 44) and star 42 in a 35-AU orbit.  The wide companion's orbit decays due to its interaction with a circumtriple disc, forcing a second exchange interaction that leaves stars 42 and 44 in a 5-AU binary and star 32 in a 32-AU orbit.  Subsequent interaction with the circumtriple disc hardens the triple and accretion hardens the close binary.  Note that continued accretion eventually results in three stellar-mass objects (Table 1).  Time is given in years after the onset of star formation.}
\end{figure}

\subsection{The formation of close binary systems}

When the calculation is stopped, amongst the 50 stars and brown 
dwarfs, there are 7 close binary systems with separations less than $10$ AU
(Table 1 and Figures \ref{pic0} and \ref{pic1}).  
Six of these close binaries are members of unstable multiple systems,
but one has been ejected from the cloud on its own and will evolve no
further (Figures \ref{pic0} and \ref{pic1}).

As described in the introduction, it seems that binaries with 
separations $\lsim 10$ AU cannot form by direct fragmentation 
because fragmentation is halted by the formation of pressure-supported 
fragments with radii of $\approx 5$ AU.  Our calculation models 
this opacity limit for fragmentation and, as expected, none of 
the fragments in our calculation form closer than $\approx 10$ AU 
from each other.  
The three smallest separations between any existing object 
and a forming fragment are 9, 21, and 22 AU and only the last of
these ends up in a close binary system.  What mechanism(s) produce 
the 7 close binary systems in our calculation?

In Figure \ref{pic1}, we trace
the histories of the stars and brown dwarfs that play a role in
producing each of the 7 close binary systems.  We find that the close 
binary systems are produced as a natural consequence of the evolution 
of wide binaries and multiple systems in a gas-rich environment.  
Three processes are involved: the accretion of gas, the interaction 
of a binary or triple with its circumbinary or circumtriple disc, 
and dynamical interactions between objects.

\subsubsection{Accretion}

Accretion aids in the production of close binaries
in two main ways.  First, accretion onto a binary
can decrease its separation (e.g.\ Figure \ref{picacc}; 
Figure 2d, steps 5--7).
This effect of gas accretion on the orbital separation of a binary 
has been studied extensively (Artymowicz 1983; Bate 1997; Bate \& Bonnell
1997; Bate 2000).  Bate \& Bonnell \shortcite{BatBon1997} show that 
accretion onto a binary from a gaseous envelope decreases the orbital 
separation of a binary unless the specific angular momentum of the 
accreted gas is significantly greater than that of the binary.  
If a binary accretes a large amount of material compared to its initial mass
(e.g.\ a factor $\gsim 10$) this can easily reduce the separation of
the binary by 1--2 orders of magnitude \cite{Bate2000}.

Second, accretion can similarly destabilise stable 
hierarchical multiple systems by reducing the separation of wide components
(e.g.\ Figure \ref{pictripacc}; Figure 2b, step 11; Figure 2c, steps 3 and 5; 
Figure 2d, step 2).  
This and other effects of accretion on the stability and mass ratios
of triple systems were studied by Smith, Bonnell \& Bate (1997).  
The destabilisation of multiple systems contributes to the 
production of close binaries because it forces the system to undergo
dynamical interactions (section \ref{dynint}).

\subsubsection{Circumbinary and circumtriple discs}

If a binary is surrounded by a circumbinary disc, gravitational torques
from the binary transfer angular momentum from the binary's orbit into
the disc, causing the binary's components to spiral together
(Artymowicz et al.\ 1991; Bate \& Bonnell 1997).  Thus, a wide
binary can be hardened into a close binary (e.g.\ Figure \ref{picacccb}; 
Figure 2b, steps 1 and 6; Figure 2d, step 7).  Similarly, a disc
around a hierarchical triple can reduce the separation of the wide
companion, destabilising the system (e.g.\ Figures \ref{pictripacc} 
and \ref{picdynacc}; Figure 2b, step 11; Figure 2c, steps 3 and 5; 
Figure 2d, step 2).  

Such disc interactions are very efficient; even relatively low-mass 
discs can have a significant effect over time \cite{Pringle1991}.  
The circumbinary and circumtriple discs in this calculation 
frequently contain a significant fraction of the system's mass
as can be observed by the spiral density waves caused
by gravitational instabilities  within them (Figures \ref{pic0}, 
\ref{picacc}, \ref{picacccb}, and \ref{picdynacc}).  This makes it
difficult to separate the effects of accretion and gravitational 
torques, but, in any case, these discs 
play an important role in reducing the separations of binaries and triples.

\subsubsection{Dynamical interactions}
\label{dynint}

Dynamical interactions can lead to the orbital evolution of a binary 
in several ways.  If the orbital velocity of a binary is greater than the
velocity of an incoming object while it is still at a great distance (i.e.\ the
binary is `hard'), the binary will survive the encounter \cite{HutBah1983}.  
However, several outcomes are possible.  The binary may simply be 
hardened by the encounter, with the single object removing energy and 
angular momentum.  Alternately, if the encounter is sufficiently close,
an unstable multiple system will be formed.  Its chaotic 
evolution will usually lead to the ejection of the object with the
lowest mass.  If the ejected object was a component of the original 
binary, the net effect is an exchange interaction.

The suggestion that dynamical encounters play an important role in 
the formation of close binaries has been made by Tokovinin (1997, 2000),
who observed that many close binaries have wider companions (see below). 
In the calculation discussed here, dynamical encounters play the 
dominant role in producing close binaries (e.g.\ Figure \ref{picdynacc}
and the exchange
interactions, hardening by fly-bys, and binary-binary interactions
described in Figure 2).  However, it is important to note that N-body
dynamical interactions alone cannot produce the high frequency
of close binaries that we obtain.  Indeed, Kroupa \& Burkert 
\shortcite{KroBur2001} find that N-body calculations which begin 
with star clusters (100 to 1000 stars) consisting entirely of 
binaries with periods $4.5 < \log{(P/{\rm days})} < 5.5$ produce 
almost no binaries with periods $\log{(P/{\rm days})}<4$.  Similarly, the
dissolution of small-N clusters typically results in binaries with 
separations only an order of magnitude smaller than the size of the initial
cluster (Sterzik \& Durisen 1998).  

The key difference in this calculation
is the effect of the gas-rich environment.  Along with the effects from
accretion and circumbinary/circumtriple discs discussed above, the
presence of gas allows dynamical encounters to be dissipative and 
to transport angular momentum.
Such dissipative encounters include star-disc encounters 
(Larson 1990; Clarke \& Pringle 1991a,b; McDonald \& Clarke 1995; 
Hall, Clarke \& Pringle 1996), where the dissipation of kinetic energy
allows the formation of bound systems from objects that would otherwise 
be unbound, and other tidal interactions (Larson 2002).

\subsection{The resulting properties of close binary systems}

As mentioned above, dynamical exchange interactions typically
lead to the ejection of the least massive component.  Thus, if a
binary encounters a star whose mass is in between the masses
of the binary's components, the binary's secondary will usually be replaced
and the binary's mass ratio will be equalised.  In addition,
gas with high specific angular momentum that accretes from
a circumbinary disc or a gaseous envelope onto a binary, 
is preferentially captured by the secondary and, thus, also drives
the mass ratio towards unity (Artymowicz 1983; Bonnell \& Bastien 1992; 
Whitworth et al.\ 1995; Bate \& Bonnell 1997; Bate 2000).
Even if accretion comes from a relatively slowly rotating cloud,
the long-term effect of this accretion is usually to equalise the
binary's components (Bate 2000). 
Thus, each of the mechanisms involved in producing close binaries
favours the production of equal-mass systems.  This is reflected 
in the mass ratios of the seven close binary systems that form in
our calculation (Table 1), all of which have values $q \gsim 0.3$ and most of
which have $q > 1/2$.
Thus, the observational result that close binaries (periods $\lsim 10$ 
years) tend to have higher mass ratios than wider binaries 
(Mazeh et al.\ 1992; Halbwachs, Mayor \& Udry 1998; Tokovinin 2000) 
is explained naturally if these processes are an integral part of 
the formation mechanism for close binaries.

Overall, the calculation produces 7 close binaries among 50 stars 
and brown dwarfs, giving a close-binary fraction of $7/43 = 16$\%.  
Given the small numbers, this is in good agreement with the observation 
that $\approx 20$\% of solar-type stars have a less massive 
main-sequence companion within 10 AU \cite{DuqMay1991}.  Thus,
dynamical encounters and orbital decay can produce the observed
frequency of close binaries.

However, because dynamical exchange
interactions preferentially select the two most massive objects,
another consequence of the way in which close binaries form 
is that the frequency of close binaries increases with the mass of 
the primary.  This can be understood by considering a binary and an
incoming object whose mass is greater than either of the two components
of the binary.  In this case, the typical result is that the incoming 
object and the primary from the original binary
form a new binary.  After several such exchange interactions, the mass 
of the primary in the new binary can be much greater than the 
original primary's mass.  In the calculation presented here, the
dependence of the frequency of close binaries on stellar mass is quite
strong.  Of $\approx 20$ brown dwarfs, there is only one
binary brown dwarf system (and if the calculation were followed further,
at least one of the components would probably accrete into the stellar 
regime), 
while of the 11 stars with masses $>0.2$ M$_\odot$, 5 are members of 
close binary systems.  While it is difficult to extrapolate from 
our low-mass star-forming cloud to larger star clusters and more
massive stars, this trend of an increasing frequency of close 
binaries with stellar mass is supported by observational surveys 
(Garmany et al.\ 1980;
Abt et al.\ 1990; Morrell \& Levato 1991; Mason et al.\ 1998).

Finally, we note that at the end of the calculation, only one of 
our close binaries has been ejected from the cloud and is on its own.
The others remain members of larger-scale bound groups
and three are members of hierarchical triple systems (Table 1; Figure 1).
This large number of wider companions
is yet another indication of the importance of multiple systems in 
producing close binaries.  
Even allowing for the eventual dissolution of the bound groups, 
it seems likely that some of the hierarchical triple systems 
will survive.  Although the true frequency of wide companions
to close binaries is not yet well known, many
close binaries do have wider components (e.g.\ Mayor \& Mazeh 1987; 
Tokovinin 1997, 2000).
Indeed, it was this observation that led Tokovinin (1997) to propose that 
dynamical interactions in multiple systems may play an important 
role in the formation of close binary systems.  Further surveys to 
determine the true frequency of wide companions to close binary 
systems would be invaluable.

\section{Conclusions}

We have presented results from a hydrodynamic calculation of the 
collapse and fragmentation of a turbulent molecular cloud to form
50 stars and brown dwarfs.  The calculation mimics the
opacity limit for fragmentation by using a barotropic equation 
of state to model the heating of collapsing gas at high densities.
This results in a minimum mass of $\approx 10$ Jupiter masses 
for the lowest-mass brown dwarfs (see Bate et al.\ 2002)
and prevents fragmentation on scales smaller than $\approx 10$ AU.
Despite this lower limit on the initial minimum separation between 
fragments, we find that as the stellar groups and unstable multiple 
systems evolve in the gas-rich environment, a high frequency of 
close binary systems (separations $\lsim 10$ AU) is produced.

Examining the history of these close binary systems, we find that they
are formed through a combination of dynamical interactions in
unstable multiple systems, and orbital decay due to accretion and/or
the interaction of binary and triple systems with circumbinary
and circumtriple discs.  These formation mechanisms allow 
realistic numbers of close binary systems to be produced without the need
for fragmentation on length scales $<10$ AU.  This avoids the difficulties
associated with the fragmentation of optically-thick gas during the 
collapse initiated by the dissociation of molecular hydrogen 
(Boss 1989; Bonnell \& Bate 1994; Bate 1998, 2002).

As a consequence of the dependence of close binary formation on dynamical 
exchange interactions and the accretion of material with high specific
angular momentum, we find that close binaries tend not to have
extreme mass ratios.  All of our systems have mass ratios $q \gsim 0.3$.
Furthermore, the frequency of close binaries is dependent on mass in
that massive stars are more likely to have close companions than lower
mass stars.
These properties are in good agreement with the results of observational 
surveys.  At the end of our calculation, many of
the close binaries are members of hierarchical triple 
systems.  Although these systems may not yet have finished evolving, the
implication is that many close binaries ought to have wider companions.
Recent observations support this hypothesis, but larger surveys to
determine the frequency of wide companions to close 
binary systems are necessary to demonstrate it conclusively.

\section*{Acknowledgments}

The computations reported here were performed using the UK Astrophysical 
Fluids Facility (UKAFF).  VB acknowledges support by the European
Community's Research Training Network under contract HPRN-CT-2000-0155,
Young Stellar Clusters.

\end{document}